\documentclass[aps,prl,twocolumn,superscriptaddress,showpacs]{revtex4}

\usepackage{graphicx}

\usepackage{hyperref}

\newcommand{\ket}[1]{\mbox{$\,\mid \! #1 \, \rangle$}}
\newcommand{\bra}[1]{\mbox{$\langle \, #1 \! \mid \,$}}

\newcommand{\cref}[1]{chapter~\ref{#1}}

\newcommand{\fref}[1]{Fig.~\ref{#1}}
\newcommand{\eref}[1]{Eq.~(\ref{#1})}

\begin{document}

%\preprint{}

\title{Experimental observation of an entire family of four-photon entangled states}

% repeat the \author  \affiliation  etc. as needed

\author{Witlef Wieczorek}
\email{witlef.wieczorek@mpq.mpg.de}
\author{Christian Schmid}
\author{Nikolai Kiesel}
\author{Reinhold Pohlner}

\affiliation{Max-Planck-Institut f{\"u}r Quantenoptik, Hans-Kopfermann-Stra{\ss}e 1, D-85748 Garching, Germany}
\affiliation{Department f\"ur Physik, Ludwig-Maximilians-Universit{\"a}t, D-80797 M{\"u}nchen, Germany}

\author{Otfried G\"uhne}
\affiliation{Institut f\"ur Quantenoptik und Quanteninformation, \"Osterreichische Akademie der Wissenschaften, A-6020 Innsbruck, Austria}
\affiliation{Institut f\"ur theoretische Physik, Universit\"at Innsbruck, Technikerstra{\ss}e 25, A-6020 Innsbruck, Austria}
\author{Harald Weinfurter}
\affiliation{Max-Planck-Institut f{\"u}r Quantenoptik, Hans-Kopfermann-Stra{\ss}e 1, D-85748 Garching, Germany}
\affiliation{Department f\"ur Physik, Ludwig-Maximilians-Universit{\"a}t, D-80797 M{\"u}nchen, Germany}


%\date{\today; version 2.2} 

\begin{abstract}
A single linear optical set-up is used to observe an entire family of four-photon entangled states. This approach breaks with the inflexibility of present linear-optical set-ups usually designed for the observation of a particular multi-partite entangled state only. The family includes several prominent entangled states that are known to be highly relevant for quantum information applications.
\end{abstract}


%\keywords{...}

\maketitle



Multi-partite entanglement is the vital resource for numerous quantum information applications like quantum computation, quantum communication and quantum metrology. So far, the biggest variety of multi-partite entangled states was studied using photonic qubits (e.g.~\cite{Bou99,Pan01,Kie07,Bou04,Xu06,Lu07}). As there is no efficient way of creating entanglement between photons by direct interaction, entangled photonic states are generally observed by a combination of a source of entangled photons and their further processing via linear optical elements and conditional detection. Based on this approach, experiments were designed for the observation of a single, e.g.~\cite{Bou99,Pan01,Kie07,Bou04,Xu06}, or two \cite{Lu07} multi-partite entangled state(s). 

Here, we break with this inflexibility by designing a single linear optics set-up for the observation of an entire family of four-photon entangled states. The states of the family are conveniently chosen by one experimental parameter. Thereby, states that differ strongly in their entanglement properties are accessible in the same experiment \cite{Wie08Note6}. We demonstrate the functionality of the scheme by the observation and analysis of a selection of distinguished entangled states.

\begin{figure}
\includegraphics[width=64mm,keepaspectratio]{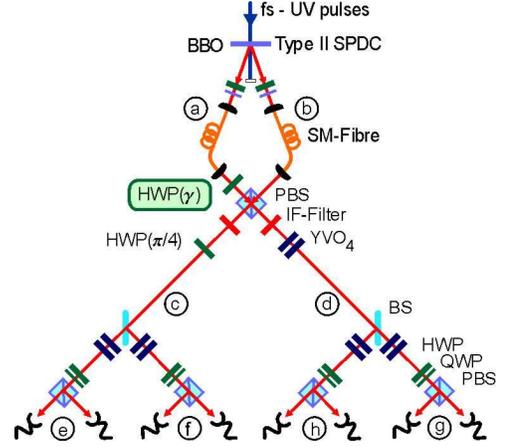}
\caption{\label{setup} Schematic experimental set-up for the observation of the family $\ket{\Psi(\gamma)}$. For details see text.}
\end{figure}

The family that can be observed experimentally is given by the superposition of the tensor product of two Bell states and a four-qubit $GHZ$ state: 
\begin{equation}
\label{eq:states}
\ket{\Psi(\gamma)}=\alpha(\gamma)\ket{\psi^+}\otimes\ket{\psi^+}+\sqrt{1-\alpha(\gamma)^2}\ket{GHZ},
\end{equation}
where $\ket{\psi^+}=1/\sqrt{2}(\ket{HV}+\ket{VH})$ and $\ket{GHZ}=1/\sqrt{2}(\ket{HHVV}+\ket{VVHH})$ \cite{Wie08Note2,Wie08Note5}. We use the notation for polarization encoded qubits, where, e.g., $\ket{HHVV}=\ket{H}_e\otimes\ket{H}_f\otimes\ket{V}_g\otimes\ket{V}_h$, and $\ket{H}$ and $\ket{V}$ denote linear horizontal and vertical polarization and the subscript denotes the spatial mode of each photon. Here, the real amplitude $\alpha(\gamma)$ with $|\alpha(\gamma)|\leq1$ is determined by a single, experimentally tunable parameter $\gamma$, which is set by the orientation of a half-wave plate (HWP). Thus, we are able to change continuously from the product of two Bell states over a number of interesting genuinely four-partite entangled states to the four-qubit $GHZ$ state. According to the four-qubit SLOCC (stochastic local operations and classical communication) classification in \cite{Ver02}, the family $\ket{\Psi(\gamma)}$ is a subset of the generic family $G_{abcd}$ of four-qubit entangled states. Note, $\ket{\Psi(\gamma)}$ represents a different class of SLOCC equivalent states for each value of $|\alpha(\gamma)|$.

The experimental set-up that allows a flexible observation of the family $\ket{\Psi(\gamma)}$ is depicted in \fref{setup}. Four photons originate from the second order emission of a spontaneous parametric down conversion (SPDC) process \cite{Kwi95} in a 2\,mm thick $\beta$-Barium borate (BBO) crystal arranged in non-collinear type II configuration. The crystal is pumped by UV pulses with a central wavelength of $390$\,nm and an average power of $600$\,mW obtained from a frequency-doubled Ti:Sapphire oscillator (pulse length $130$\,fs). The four photons are emitted into two spatial modes $a$ and $b$ \cite{Wei01}: 
\begin{equation}
\label{eq:spdc}
1/(2\sqrt{3})[(a_H^\dagger b_V^\dagger)^2+(a_V^\dagger b_H^\dagger)^2+2a_H^\dagger a_V^\dagger b_H^\dagger b_V^\dagger]\ket{\mathrm{vac}},
\end{equation}
where $m_j^\dagger$ is the creation operator of a photon having polarization $j$ in mode $m$ and $\ket{\mathrm{vac}}$ is the vacuum state. A HWP and a 1\,mm thick BBO crystal compensate walk-off effects. The spatial modes $a$ and $b$ are defined by coupling the photons into single mode (SM) fibres. Spectral selection is achieved by $3$\,nm FWHM interference filters (IF) centered around $780$\,nm. A HWP in mode $a$ transforms the polarization of the photons. The orientation of the optical axis, $\gamma$, of this HWP is the tuning parameter of the family. Subsequently, the modes $a$ and $b$ are overlapped at a polarizing beam splitter (PBS) with its output modes denoted by $c$ and $d$. A HWP oriented at $\pi/4$ behind the PBS transforms the polarization of the photons in mode $c$ from $H(V)$ into $V(H)$. Subsequently, the modes $c$ and $d$ are split into the output modes $e,f$ and $g,h$, respectively, via polarization-independent beam splitters (BS). Birefringence of the beam splitters is compensated by a pair of perpendicularly oriented birefringent Yttrium-Vanadate (YVO$_4$) crystals. Finally, the polarization state of each photon is analyzed with a HWP, a quarter-wave plate (QWP) and a PBS. The photons are detected by fiber-coupled single photon detectors and registered by a multichannel coincidence unit. 

\begin{figure}
\includegraphics[width=76mm,keepaspectratio]{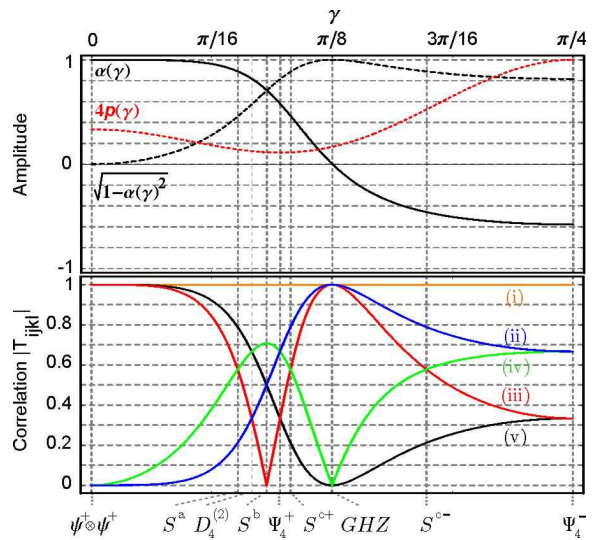}
\caption{\label{fig:amplcorrs} The upper panel shows the dependence of the amplitudes $\alpha(\gamma)$ (solid) and $\sqrt{1-\alpha(\gamma)^2}$ (dashed) on the tunable parameter $\gamma$ for the family $\ket{\Psi(\gamma)}$. Also the probability $p(\gamma)$ (dotted) to observe the states $\ket{\Psi(\gamma)}$ is shown. The lower panel shows the modulus of the correlations, $|T_{ijkl}|$, for the family $\ket{\Psi(\gamma)}$: (i) $T_{iiii}$ with $i\in\{0,x,y,z\}$; (ii) $T_{0z0z}$, $T_{xyxy}$; (iii) $T_{00zz}$, $T_{xxyy}$; (iv) $T_{ijij}$ and (v) $T_{iijj}$ with $i\in\{0,z\}$, $j\in\{x,y\}$. In order to obtain all 40 correlations, following permutations starting from a normal ordering $(1,2,3,4)$ are necessary: $(1,2)\leftrightarrow (3,4)$, $(1)\leftrightarrow (2)$ and $(3)\leftrightarrow (4)$.}
\end{figure}

Under the condition of detecting one photon of the second order SPDC emission in each spatial output mode the family of states $\ket{\Psi(\gamma)}$ is observed, where the amplitude $\alpha(\gamma)$ depends on the HWP angle $\gamma$ via $\alpha(\gamma)=(2\cos{4\gamma})/\sqrt{48p(\gamma)}$ with $\gamma\in[0,\frac{\pi}{4}]$. This occurs with a probability $p(\gamma)=(5-4\cos{4\gamma}+3\cos{8\gamma})/48$ (\fref{fig:amplcorrs}). Only for few states of the family a dedicated set-up is known \cite{Pan01,Kie07,Bou04,Xu06}. For these particular cases the respective state is observed with equal or higher probability. Here, however, we profit from the flexibility to choose various entangled states using the same set-up. 

Let us illustrate the described state observation scheme by examining the action of the HWP together with the PBS. We note that only the case where two photons are found in each spatial mode $c$ and $d$ behind the PBS, respectively, can lead to a detection event in each of the four output modes $e,f,g,h$. First, we consider a HWP oriented at $\gamma=0$. This setting leaves the polarization of each photon unchanged. Each of the first two terms of \eref{eq:spdc} results in four photons in the same spatial mode behind the PBS and, thus, does not contribute to a fourfold coincidence in the output modes. However, the last term of \eref{eq:spdc} yields two photons in each mode behind the PBS, whose state is $\propto c_H^\dagger c_V^\dagger d_H^\dagger d_V^\dagger$. A symmetric distribution of these photons leads to the observation of a Bell state in modes $e,f$ and in modes $g,h$, respectively: $\ket{\Psi(0)}=\ket{\psi^+}\otimes\ket{\psi^+}$. Conversely, the last term of \eref{eq:spdc} can be suppressed by interference when the HWP is oriented at $\gamma=\pi/8$ transforming $H/V$ into $+/-$ polarization [$\ket{\pm}=1/\sqrt{2}(\ket{H}\pm\ket{V})$]. Then, two photons in each mode $c$ and $d$ can only originate from the first two terms of \eref{eq:spdc} and result in the state $\propto (c_H^\dagger d_H^\dagger )^2+(c_V^\dagger d_V^\dagger )^2$ directly behind the PBS. This yields the $GHZ$ state in the output modes. Continuous tuning of the HWP in the range $\gamma\in(0,\pi/8)$ and $\gamma\in(\pi/8,\pi/4)$ leads to any superposition of the states $\ket{\psi^+}\otimes\ket{\psi^+}$ and $\ket{GHZ}$ and, thus, to the observation of the entire family of states. 

This family contains useful states, which, moreover, differ strongly in their entanglement properties. For example, the well known $GHZ$ state [$\ket{GHZ}=\ket{\Psi(\pi/8)}$, i.e.~$\alpha=0$] \cite{Pan01} belongs to the graph states \cite{Hei04} and finds numerous applications in quantum information, e.g.~\cite{Hil99}. The entanglement of the symmetric Dicke states \cite{Dic54} is known to be very robust against photon loss. Out of these states we observe with $\alpha=\sqrt{2/3}$ the state $\ket{D_4^{(2)}}=\ket{\Psi(\pi/12)}$ \cite{Kie07}. Remarkably, this state allows to obtain, via a single projective measurement, states out of each of the two inequivalent classes of genuine tri-partite entanglement \cite{Kie07,Dur00}. The states $\ket{\Psi_4^-}=\ket{\Psi(\pi/4)}$ ($\alpha=-\sqrt{1/3}$) \cite{Bou04} and $\ket{\psi^-}\otimes\ket{\psi^-}$ \cite{Bou04DF} [that are equivalent under local unitary (LU) operations to $\ket{\Psi_4^+}=\ket{\Psi(\approx0.098\pi)}$ ($\alpha=\sqrt{1/3}$) \cite{Xu06} and $\ket{\psi^+}\otimes\ket{\psi^+}=\ket{\Psi(0)}$ ($\alpha=1$), respectively] are invariant under any action of the same LU transformation on each qubit and, therefore, they form a basis for decoherence-free communication \cite{Kem01}. 

To characterize the family of states we consider the correlations of $\ket{\Psi(\gamma)}$. Out of all 256 correlations $T_{ijkl}$ \cite{Wie08Note3} in the standard basis, the family $\ket{\Psi(\gamma)}$ exhibits at most 40 that are non-zero. The modulus of these correlations, $|T_{ijkl}|$, shows five distinct dependencies on $\gamma$, which are shown in \fref{fig:amplcorrs}. Interestingly, one finds the aforementioned states at the crossing points of some correlations. Consequently, we can identify other distinguished states at the remaining four crossing points. These are found at $\gamma\approx0.076\pi$ [$\alpha=(1/6(3+\sqrt{3}))^{1/2}$], $\gamma\approx0.091\pi$ ($\alpha=\sqrt{1/2}$), $\gamma\approx0.1034\pi$ [$\alpha=(1/6(3-\sqrt{3}))^{1/2}$], and for $\gamma\approx0.174\pi$ [$\alpha=-(1/6(3-\sqrt{3}))^{1/2}$]. 
We label them for brevity by $\ket{S^a}$, $\ket{S^b}$, $\ket{S^{c+}}$ and $\ket{S^{c-}}$, respectively.

\begin{figure}
\includegraphics[width=80mm,keepaspectratio]{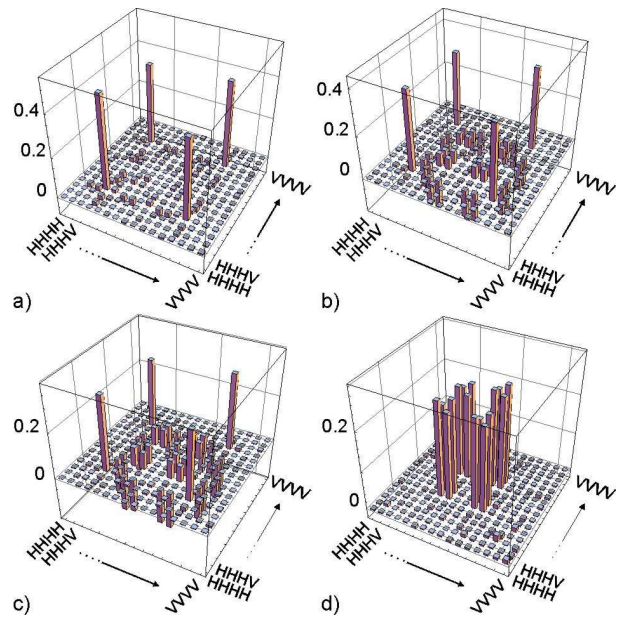}
\caption{\label{fig:densitymatrices} Real part of experimental density matrices for the states (a) $\ket{GHZ}$, (b) $\ket{S^{c-}}$, (c) $\ket{\Psi_4^-}$ and (d) $\ket{\psi^+}\otimes\ket{\psi^+}$. For the states $\ket{\Psi_4^-}$ and $\ket{GHZ}$ the imaginary part has a peak at the off-diagonal element $\ket{HHHH}\bra{VVVV}$ of 0.06 and 0.08, respectively, representing a slight imaginary phase between the terms $\ket{HHHH}$ and $\ket{VVVV}$. Otherwise noise on the real and imaginary part is comparable.}
\end{figure}

We select these nine states for an experimental characterization. As the set-up is stable and delivers the states with a reasonable count rate we are able to perform state tomography on $\ket{GHZ}$, $\ket{S^{c-}}$, $\ket{\Psi_4^-}$ and $\ket{\psi^+}\otimes\ket{\psi^+}$ of the selected set. The full tomographic data set was obtained from 81 different analysis settings for each state \cite{Kie07}. Due to the different probabilities to observe these states we varied the total measurement time between 54 hours for $\ket{\Psi_4^-}$ and 202.5 hours for $\ket{GHZ}$ with count rates of $23.2$\,min$^{-1}$ and $4.9$\,min$^{-1}$, respectively, without any realignment during each measurement run. The resulting density matrices are displayed in \fref{fig:densitymatrices}. The population and coherence terms for a $GHZ$ state are clearly visible in \fref{fig:densitymatrices}(a). In \fref{fig:densitymatrices}(b) additional to the $GHZ$ part the population and coherence terms of the $\ket{\psi^+}\otimes\ket{\psi^+}$ component appear. The (negative) coherence terms show that indeed a coherent superposition of both parts is achieved. The same structure is visible in \fref{fig:densitymatrices}(c) with an increased $\ket{\psi^+}\otimes\ket{\psi^+}$ part. Finally, in \fref{fig:densitymatrices}(d), the $GHZ$ part has disappeared completely. This clearly illustrates that we are able to tune the relative weight between the states $\ket{\psi^+}\otimes\ket{\psi^+}$ and $\ket{GHZ}$ coherently, instead of only mixing them.

\begin{figure}
\includegraphics[width=76mm,keepaspectratio]{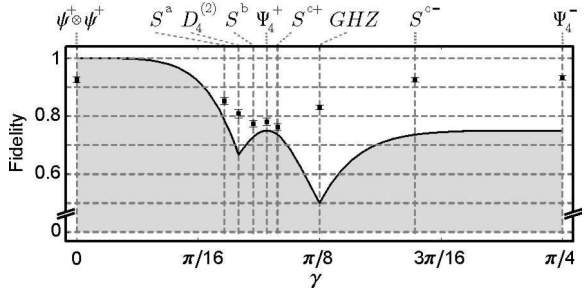}
\caption{\label{fig:results} Experimentally determined fidelities of nine distinguished states from the family $\ket{\Psi(\gamma)}=\alpha(\gamma)\ket{\psi^+}\otimes\ket{\psi^+}+\sqrt{1-\alpha(\gamma)^2}\ket{GHZ}$. The minimal fidelity for proving genuine four-qubit entanglement is depicted as solid curve.}
\end{figure}

Next, we focus on the quality of the states and on proving their entanglement. As a measure of the former we evaluate the fidelity $F_{\Psi(\gamma)}=\bra{\Psi(\gamma)}\rho_{\mathrm{exp}}\ket{\Psi(\gamma)}$ for the observed states $\rho_{\mathrm{exp}}$, where at most 21 measurement settings are required for the determination of $F_{\Psi(\gamma)}$ \cite{Wie08Note4}. To perform these measurements for the remaining five states the total measurement time ranged from 45.5 hours for $\ket{S^{a}}$ up to 112 hours for $\ket{\Psi_4^{+}}$, with count rates of $4.1$\,min$^{-1}$ and $1.6$\,min$^{-1}$, respectively. The fidelities for all states are depicted in \fref{fig:results}. We find high fidelities ranging from 0.75 up to 0.93. Obviously, the fidelity shows a dependence on $\gamma$. We emphasize that this behavior is not caused by a different optical alignment for each state, rather, it can be qualitatively attributed to different effects. Higher order emissions of the SPDC, which can lead to additional four-fold coincidences, reduce the fidelity. For the actual experimental parameters (pair generation probability, coupling and detection efficiencies) we calculated that the fidelity for $\gamma=0,\pi/4$ would be reduced by about $1\%$, while a reduction of up to $8\%$ would be found for states around $\ket{\Psi_4^{+}}$. Furthermore, the fidelity of the observed states relies on the indistinguishability of the SPDC photons \cite{Gri97} and on the quality of interference. While for $\gamma=0,\pi/4$ the PBS acts in the computational basis as a polarization filter only, for all other $\gamma$ imperfect interference is relevant \cite{Wie08Note7} and, thus, leads to an additional reduction of the fidelity. 
%This applies especially to states around $\ket{\Psi_4^{+}}$ as seen in \fref{fig:results}. 
Considering these effects, the question arises whether the fidelity of particular states is higher when these states were observed with dedicated linear optics set-ups. For example, the states $\ket{D_4^{(2)}}$ and $\ket{\Psi_4^{-}}$ were recently observed with fidelities of $F_{D_4^{(2)}}=0.844\pm0.008$ \cite{Kie07} and $F_{\Psi_4^-}=0.901\pm0.01$ \cite{Bou04}, respectively. Here we achieved $0.809\pm0.014$ and $0.932\pm0.008$, respectively, comparable with the dedicated implementations.

Finally, for proving genuine four-partite entanglement of the observed states we apply generic entanglement witnesses $\mathcal{W}_{\Psi(\gamma)}$ \cite{Bou04,Hor96}. Their expectation value depends directly on the fidelity: $\mathrm{Tr}(\mathcal{W}_{\Psi(\gamma)}\rho_{\mathrm{exp}})=c(\gamma)-F_{\Psi(\gamma)}$, where $c(\gamma)$ is the maximal overlap of $\ket{\Psi(\gamma)}$ with all bi-separable states. A fidelity larger than $c(\gamma)$ (solid curve in \fref{fig:results}) detects genuine four-qubit entanglement of $\rho_{\mathrm{exp}}$. We find that all experimental fidelities, of course except $F_{\Psi(0)}$, are larger than $c(\gamma)$, thus, proving four-qubit entanglement. For the bi-separable entangled state $\ket{\Psi(0)}$ we apply the witness given in \cite{Guh02} on each pair and find $-0.466\pm0.006$ and $-0.461\pm0.006$, respectively, detecting the entanglement of each pair.

To summarize, we are able to observe an entire family of highly entangled four-photon states with high fidelity by using the same linear optics set-up. For this purpose, a single SPDC source and one overlap on a PBS were sufficient. This is a clear improvement compared to previous dedicated linear optics realizations, where basically only one state could be observed. The general principle of commonly manipulating multi-photon states followed by interferometric overlaps at linear optical components, of course, can be easily extended: For example, one can use the six photon emission from the SPDC source and the presented set-up, or replace the PBS with a BS. Both enables the observation of different families of states \cite{Wie08Note1}. Even if the weak photon-photon coupling does not allow the design of simple quantum logic gates, the utilization of higher order emissions from an SPDC source together with multi-photon interference will enable further flexible experiments, each with numerous different and highly relevant multi-partite entangled states.
\begin{acknowledgments}
We thank W.~Laskowski and M.~Bourennane for stimulating discussions. We acknowledge the support of this work by the DFG-Cluster of Excellence MAP, the FWF and the EU Projects QAP, SCALA, QICS and OLAQUI. W.W.~acknowledges support by QCCC of the ENB and the Studienstiftung des dt.~Volkes.
\end{acknowledgments}

%\bibliographystyle{h-physrev}
%\bibliography{citations}

\begin{thebibliography}{10}
\bibitem{Wie08Note2}
For each $\ket{\Psi(\gamma)}$ a variety of other states can be obtained by
  local operations, which are easily implementable by linear optical elements.
  While we focus on polarization encoding, the principle of state
  manipulations described here can be applied for other qubit encodings, too.
\bibitem{Wie08Note3}
The correlations $T_{ijkl}$ of a four qubit pure state $\ket{\psi}$ are given
  by $T_{ijkl}=\bra{\psi}
  (\sigma^i\otimes\sigma^j\otimes\sigma^k\otimes\sigma^l) \ket{\psi}$, where
  $i,j,k,l\in\{0,x,y,z\}$, $\sigma^0 = \openone$ and $\sigma^{x,y,z}$ are the
  usual Pauli operators.
\bibitem{Wie08Note4}
When writing $\rho_{\mathrm{exp}}$ in the basis of the Pauli operators $\rho_{\mathrm{exp}}=
  \sum_{ijkl}
  S_{ijkl}(\sigma^i\otimes\sigma^j\otimes\sigma^k\otimes\sigma^l)/16$, where
  $S_{ijkl}$ are the correlations of the experimentally observed state, we
  find $F_{\Psi(\gamma)} = \sum_{ijkl}S_{ijkl}T_{ijkl}/16$. Only terms with
  non-zero $T_{ijkl}$ contribute to $F_{\Psi(\gamma)} $.
\bibitem{Wie08Note7}
We find a minimal two-photon visibility at the PBS for $\gamma\approx 1.2\pi/16$ of $(94.6\pm1.5)\%$.
\bibitem{Wie08Note1}
For example, the third order emission of the SPDC is overlapped on the PBS. Subsequently, the photons are distributed onto six spatial modes. Then, the HWP($\gamma$) in front of the PBS tunes the observable six photon states: $\beta(\gamma) \ket{GHZ_6} + \sqrt{1-\beta(\gamma)^2}/\sqrt{2} (\ket{\overline{W}_3}\otimes\ket{{W}_3} - \ket{W_3}\otimes\ket{\overline{W}_3})$, where $\ket{GHZ_6}=1/\sqrt{2}(\ket{HHHVVV}-\ket{VVVHHH})$, $\ket{W_3}=1/\sqrt{3}(\ket{HHV}+\ket{HVH}+\ket{VHH})$, $\ket{\overline{W}_3} = 1/\sqrt{3}(\ket{VVH}+\ket{VHV}+\ket{HVV})$.
\end{thebibliography}

\vspace{\baselineskip}
\copyright \, 2008 The American Physical Society

\end{document}